\documentclass[aps,pra,showpacs,twocolumn]{revtex4}

\usepackage{dcolumn}
\usepackage{bm}
\usepackage{amsmath}
\usepackage{graphicx}
\usepackage{amsmath,amssymb}

\newcommand{\bra}[1]{\langle{#1}|}
\newcommand{\ket}[1]{|{#1}\rangle}

\usepackage{graphics} 
\usepackage{epsfig} 
\usepackage{times} 
\usepackage{amsmath} 
\usepackage{amssymb}  

\begin{document}

\title{Distributed entanglement generation between continuous-mode 
Gaussian fields with measurement-feedback enhancement}



\author{Hendra I. Nurdin\footnote{School of Electrical Engineering and Telecommunications,  
The University of New South Wales,  
Sydney NSW 2052, Australia; h.nurdin@unsw.edu.au}}
 \affiliation{%
Research School of Engineering, The Australian National University, Canberra ACT 0200, Australia.
}

\author{Naoki Yamamoto\footnote{yamamoto@appi.keio.ac.jp}}
 \affiliation{%
Department of Applied Physics and Physico-Informatics,
Keio University, 
Hiyoshi 3-14-1, Kohoku-ku, Yokohama 223-8522, Japan.
}


\date{\today}


\begin{abstract} 

This paper studies a scheme of two spatially distant oscillator 
systems that are connected by Gaussian fields and examines 
distributed entanglement generation between two continuous-mode 
output Gaussian fields that are radiated by the oscillators. 
It is demonstrated that using measurement-feedback control 
while a non-local effective entangling operation is on can help 
to enhance the Einstein-Podolski-Rosen (EPR)-like entanglement 
between the output fields. 
The effect of propagation delays and losses in the fields 
interconnecting the two oscillators, and the effect of other 
losses in the system, are also considered. 
In particular, for a range of time delays the measurement feedback 
controller is able to maintain stability of the closed-loop 
system and the entanglement enhancement, but the achievable 
enhancement is only over a smaller bandwidth that is commensurate 
with the length of the time delays.

\end{abstract}


\pacs{03.67.Bg, 02.30.Yy, 42.50.Dv}




\begin{abstract}

This paper studies a scheme of two spatially distant oscillator 
systems that are connected by Gaussian fields and examines 
distributed entanglement generation between two continuous-mode 
output Gaussian fields that are radiated by the oscillators. 
It is demonstrated that using measurement-feedback control 
while a non-local effective entangling operation is on can help 
to enhance the Einstein-Podolski-Rosen (EPR)-like entanglement 
between the output fields. 
The effect of propagation delays and losses in the fields 
interconnecting the two oscillators, and the effect of other 
losses in the system, are also considered. 
In particular, for a range of time delays the measurement feedback 
controller is able to maintain stability of the closed-loop 
system and the entanglement enhancement, but the achievable 
enhancement is only over a smaller bandwidth that is commensurate 
with the length of the time delays.

\end{abstract}


\maketitle

\newtheorem{thm}{Theorem}
\newtheorem{lemma}{Lemma}
\newtheorem{prop}{Proposition}
\newtheorem{deff}{Definition}
\newtheorem{remark}{Remark}


\maketitle
\thispagestyle{empty}
\pagestyle{empty}

\section{Introduction} 

Entanglement between quantum systems is considered to be an 
important resource to be exploited in many quantum-based 
technologies proposed in recent years. 
In particular, effective entanglement distribution in a 
quantum network is a problem that has attracted attention 
in the literature due to its importance in applications 
\cite{Cirac1997,Kimble2007}. 
However, entanglement is very fragile and the amount of 
entanglement can quickly be lost due to decoherence. 
One strategy to overcome this is to use several copies of 
quantum systems with a limited degree of entanglement 
and to process them to obtain a single copy that contains 
a higher degree of entanglement. 
This process is known as entanglement distillation 
\cite{Bennett1996}. 

Researchers have considered entanglement distillation in 
both discrete and continuous variables. 
In the continuous variable case, a particular class of 
systems of interest are oscillator systems that are in 
a Gaussian state \cite{Ferraro2005}. 
If one has several copies of bipartite entangled pairs of 
Gaussian states then a no-go result of \cite{Eisert2002} 
states that it is not possible to distill further entanglement 
using only Gaussian local operation and classical communication 
(LOCC). 
Essentially, because the operations are LOCC there can be 
no entangling operations between any of the oscillators as 
they would necessarily have to be non-local. 
However, often in practice one considers dynamical quantum 
systems having an entangling interaction on during the evolution. 
The effectiveness of entanglement generation may be hampered 
by the decoherence taking place, possibly causing the 
entanglement to eventually vanish. 
Since there is a non-local entangling operation in effect, 
the no-go theorem does not hold, and this opens possibility 
to use Gaussian LOCC operations to protect entanglement during 
the system evolution. 
Thus, Mancini and Wiseman \cite{Mancini2007} considered 
the use of measurement feedback to improve entanglement between 
 two cavity modes $a_1$ and $a_2$ coupled via 
a two mode squeezing Hamiltonian $H = i(\epsilon a_1^*a_2^*-\epsilon^* a_1a_2)$; in this paper the notation $^*$ denotes the the adjoint of a Hilbert space operator or the conjugate transpose of a matrix of complex numbers or operators. At the same time, the two modes undergo decoherence, due to the cavity photons escaping through the transmissive cavity mirrors on which measurements can be made.  
They showed that there exists a Gaussian LOCC strategy 
realized by measurement-feedback that can help to improve 
the amount of entanglement between the two cavity modes 
compared to when no measurement and feedback is used. 
It was subsequently shown in \cite{Serafini2010} 
that the strategies proposed in \cite{Mancini2007} 
are optimal for that particular physical setup. 
Several other works in the literature have also considered 
using feedback for entanglement control, e.g., 
\cite{Carvalho2008,Yamamoto2008,WisemanBook,Yan2011}.

The main contribution of this paper is developing a feedback-controlled 
scheme that uses distributed resources (parametric amplification at two spatially separate sites) to generate entanglement between two spatially separated continous-mode Gaussian fields. The system of interest is a quantum network consisting of two  spatially separated open Gaussian oscillator systems. 
They are interconnected via travelling Gaussian quantum fields 
that act as common baths between the two oscillators and as 
a source of effective interaction between them. The scheme also exploits measurement-feedback control
to enhance the entanglement. It is distinct from a measurement-feedback control based scheme on a linear quantum system studied in the earlier work  \cite{Mancini2007} in several ways:
\begin{enumerate}

\item The scheme uses distributed resources. 
That is, entanglement is generated exploiting
contributed resources at two spatially separated locations 
(say, at Alice and Bob's). 
This use of spatially distributed resources for entanglement 
generation is a key feature of the proposed scheme. 
In contrast, \cite{Mancini2007} considers two oscillators 
interacting in a  $\chi^{(2)}$ nonlinear crystal at a single 
location, and the entangling process occurs only at one site 
using only resources available at that site.

\item The paper is concerned with entanglement between two output fields, that consist of a continuum of modes, rather than entanglement between single mode internal
oscillator modes as in \cite{Mancini2007}. Although the system in \cite{Mancini2007} does have output fields, they are  only provided as observables to be measured and fedback rather than left freely as entangled resources.
In contrast, the present work utilizes  several output fields, some of which are  for measurement and feedback while others remain free to be used as entangled resources. 
The rationale for considering entanglement in the output fields, rather than between internal oscillator modes, is that the output fields are more easily accessible to be exploited for various purposes.  For instance, if multiple copies of the system are available then the multiple outputs of the various copies could, say, be passed onto some (at this stage hypothetical) entanglement distiller to produce  another signal with improved EPR-like qualities.

\item Since generation of entanglement is via  transmission channels linking the two spatially separated sites, there are inherent features of this  scheme that need to be considered and which are not present in [6]. The first is the presence of losses along these transmission channels, and the second is the presence of time delays required for the interconnecting fields to propagate between the two sites.  
\end{enumerate}

In this work we show that a {\it linear quadratic Gaussian} (LQG) 
measurement-feedback controller 
\cite{WisemanBook,Doherty1999,Belavkin2008} can enhance 
the entanglement between the output fields as compared to when there 
is no controller present. 
Moreover, this controller can provide an enhancement 
even in the presence of losses and delays in the overall system. That is,
the controller displays robustness with respect to the presence of these imperfections.
Simulation results are presented and discussed to compare the 
performance of the uncontrolled and controlled systems in the ideal 
case and in the presence of imperfections. 


\section{Preliminaries}
\label{sec:prelim}

\subsection{Linear quantum systems }
\label{sec:linear-qsys}

In general, the dynamics of an open quantum system, which does 
not contain a scattering process, can be characterized by the 
system-environment coupling Hamiltonian 
\begin{equation}
\label{interaction}
   H_{\rm int}(t) = i\sum_{j=1}^m (L_j\xi_j(t)^* - L_j^* \xi_j(t)), 
\end{equation}
where $\xi_j(t)~(j=1,\ldots,m)$ is the field operator describing 
the $j$-th environment field and $L_j$ is the system operator 
corresponding to the $j$-th coupling \cite{GardinerBook}. 
When the Markov limit is imposed on the environment, the field 
operators satisfy 
$[\xi_i(t), \xi_j(s)^*]=\delta(t-s) \delta_{ij}$. 

A linear quantum system with $n$-bosonic modes $a_j(t)~(j=1,\ldots, n)$ 
satisfying $[a_i, a_j^*]=\delta_{ij}$ appears when $L_j$ is linear and 
$H$ is quadratic in $a_j$ and $a_j^*$. 
In this case, the Heisenberg equation of $a_j(t)=U(t)^* a_j U(t)$ 
with unitary 
$U(t)={\rm exp}^{\hspace{-0.5cm}\longrightarrow}~
(-i\int_0^t H_{\rm int}(s)ds)$ has the following form: 
\begin{equation}
\label{dynamics}
    \dot{z}(t)=Az(t)+B\xi(t), 
\end{equation}
where we have defined 
\begin{equation*}
    z=(a_1^q, a_1^p, \ldots, a_n^q, a_n^p)^T,~~
    \xi=(\xi_1^q, \xi_1^p, \ldots, \xi_m^q, \xi_m^p)^T,
\end{equation*}
with {\it quadratures} $a_j^q=a_j+a_j^*$, $a_j^p=(a_j-a_j^*)/i$, 
$\xi_j^q=\xi_j+\xi_j^*$, and $\xi_j^p=(\xi_j-\xi_j^*)/i$. 
The field operator also changes to 
$\xi_{out,j}(t)=U(t)^*\xi_j(t)U(t)$, and measuring this output 
field generates the classical signal
\begin{equation}
\label{output}
    y(t)=Cz(t)+D\xi(t). 
\end{equation}
The system matrices $A, B, C$, and $D$ have specific structures 
to satisfy $[a_i(t), a_j(t)^*]=\delta_{ij}$ for all $t \geq 0$, and $[y(t),y(s)]=0$ for all $s,t \geq 0$. 


\subsection{Measurement-feedback LQG control}
\label{sec:lqg-measurement-feedback}

The field $\xi(t)$ can closely be approximated by a coherent 
light field generated from a laser device. 
Thus feedback control can be realized by replacing $\xi_j(t)$ 
by a modulated field $u_j(t)+\xi_j(t)$, 
where $u_j(t)$ is a function of $y(s)~(s\leq t)$. 

In the LQG feedback control scheme, the control input 
$u(t)=(u_1^q, u_1^p, \ldots, u_m^q, u_m^p)$ is generated as 
an output of the following classical linear system with $y(t)$ as the input: 
\begin{equation}
\label{controller}
   \dot{z}_c(t)=A_c z_c(t) + B_c y(t),~~~
   u(t)=C_c z_c(t). 
\end{equation}
$z_c(t)$ is a vector of real c-numbers representing the state 
of controller. 
$A_c,~B_c$, and $C_c$ are real matrices to be designed. 
Combining Eqs. \eqref{dynamics}, \eqref{output}, and 
\eqref{controller}, we have a closed-loop dynamics with 
variable $\tilde{z}=(z^T,z_c^T)^T$. 
For this system, we consider the following quadratic-type cost 
function: 
\begin{equation}
\label{LQG cost}
    J(u) 
      = \mathop{\lim}_{T \rightarrow \infty} \frac{1}{T}
            {\mathbb E}\Big[
                \int_{0}^T \Big\{ 
                   \tilde{z}(t)^T P \tilde{z}(t) 
                     + u(t)^T Q u(t) 
                    \Big\} dt \Big], 
\end{equation}
where $P\geq 0$ and $Q>0$ are weighting matrices that should be 
chosen appropriately. 
The expectation is defined as 
${\mathbb E}(X)={\rm Tr}[X(\rho\otimes\ket{0}\bra{0})]$, 
where $\rho$ is the initial system Gaussian state and $\ket{0}$ is the 
field vacuum state. 
The optimal LQG feedback control input is given as a minimizer 
of the cost function \eqref{LQG cost}; 
the minimization problem $\min J(u)$ can be efficiently solved 
using the Matlab Control System Toolbox to yield the optimal 
set of matrices $A_c, B_c$, and $C_c$. 


\subsection{Frequency domain entanglement criteria for 
continuous-mode output fields}
\label{sec:entanglement-criterion}

Let us here particularly focus on two output fields 
$\xi_{out,1}(t)$ and $\xi_{out,2}(t)$. 
These are continuous-mode fields, therefore we need to 
move to the frequency domain to evaluate their entanglement; 
for the quadratures $x_j=\xi_{out,j}+\xi_{out,j}^*$ 
and $y_j=(\xi_{out,j}-\xi_{out,j}^*)/i$, define their Fourier 
transforms by 
$O_j(i\omega)=\int o_j(t)e^{-i\omega t} dt/\sqrt{2\pi}$ 
with $o=x, y$ and $O=X, Y$, respectively. 
Then the two output fields $\xi_{out,1}(t)$ and $\xi_{out,2}(t)$ 
are entangled for the mode at frequency $\omega$ if \cite{Vitali2006}
\begin{equation}
\label{eq:entangle-criterion}
   V_+(i\omega) + V_-(i\omega)<4,
\end{equation}
where $V_+(i\omega)$ and $V_-(i\omega)$ are defined by the identities 
\begin{align*}
   \lefteqn{\langle(X_1(i\omega) + X_2(i\omega))^*
       (X_1(i\omega') + X_2(i\omega'))\rangle }\\ 
   &~~ = \langle(X_1(-i\omega) + X_2(-i\omega))
             (X_1(i\omega') + X_2(i\omega'))\rangle \\
   &~~ = V_+(i\omega)\delta(\omega-\omega'), \\
   & \lefteqn{\langle (Y_1(i\omega) - Y_2(i\omega))^*
             (Y_1(i\omega') - Y_2(i\omega'))\rangle} \\
   &~~ =\langle (Y_1(-i\omega) - Y_2(-i\omega))
             (Y_1(i\omega') - Y_2(i\omega'))\rangle \\  
   &~~ = V_-(i\omega)\delta(\omega-\omega'). 
\end{align*}
That is, $V_+(i\omega)$ and $V_-(i\omega)$ are power spectral 
densities of the fields. 
The inequality \eqref{eq:entangle-criterion} is a well-known sufficient 
condition for entanglement in the frequency domain \cite{Ou1992,Vitali2006}. 
In the case of a two mode-squeezed state, the two power spectra 
are identical and, in an ideal limit, they converge to zero for 
all $\omega$, implying that the so-called Einstein-Podolski-Rosen 
(EPR) pair of fields is produced \cite{Ou1992}. 


\section{The system model} 
\label{sec:sys-model}

\subsection{Description}

\begin{figure*}[tbph]
\centering
\includegraphics[scale=0.65]{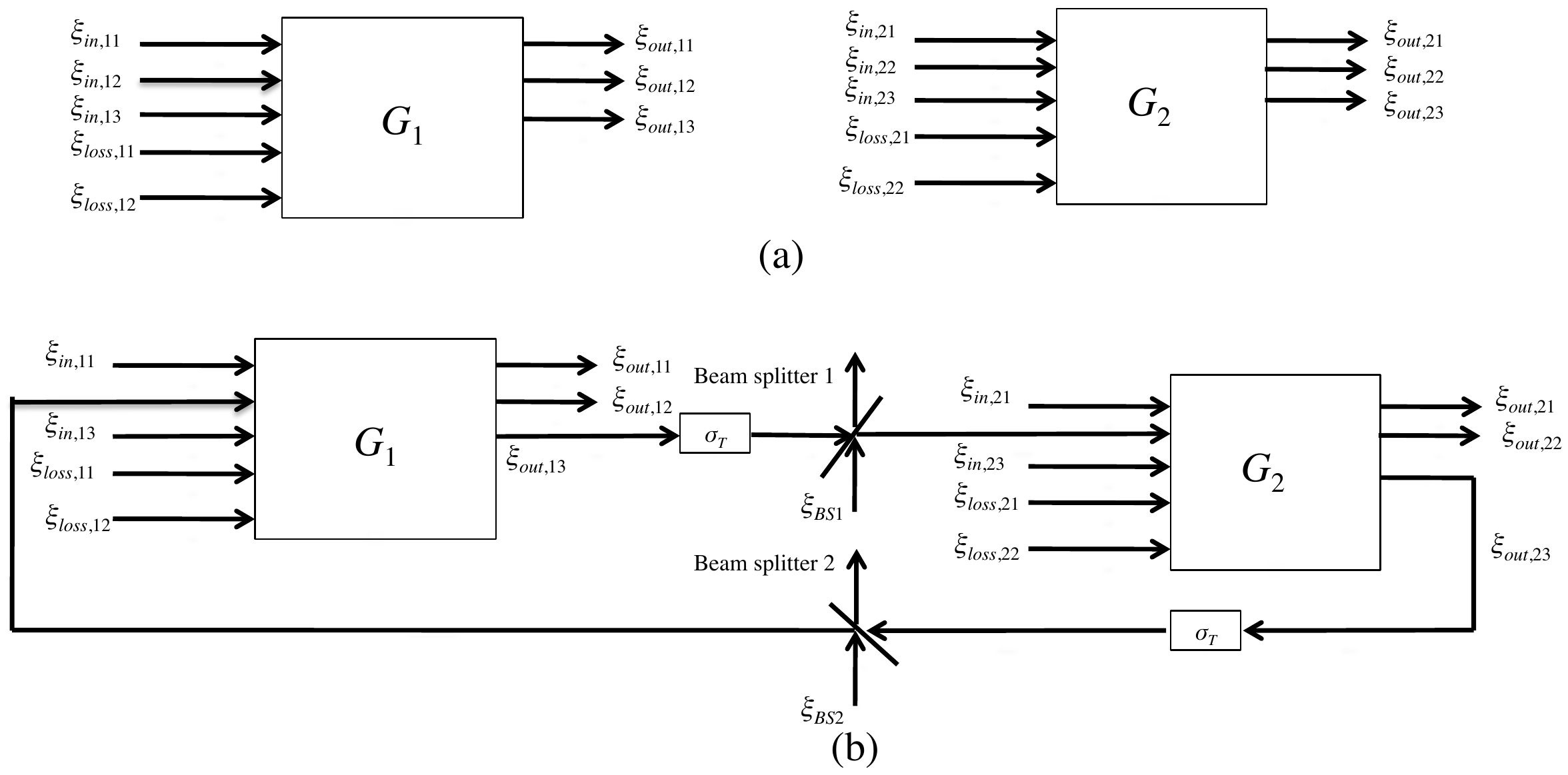}
\caption{
(a) Input and output fields of the open oscillators systems 
$G_1$ and $G_2$. 
In both systems, the output fields corresponding to the loss 
field $\xi_{loss,ij}$ cannot be essentially observed, hence 
they are not shown. 
(b) The connected open oscillator systems of $G_1$ and $G_2$. 
The traveling fields contain possible losses modeled by two beam 
splitters BS$1$ and BS$2$ with vacuum noises $\xi_{BS,1}$ and 
$\xi_{BS,2}$ entering into their unused ports, respectively. 
$\sigma_T$ denotes the operation bringing a time-delay $T$ on 
the traveling fields. 
}
\label{fig:nocontrol_net}
\end{figure*}

We consider the system shown in Fig.~\ref{fig:nocontrol_net}. 
The spatially separated open quantum systems $G_1$ and $G_2$ 
are connected by Gaussian quantum fields. 
The system $G_j$ consists of two oscillator modes $a_j$ and $b_j$ 
having the same oscillation frequency coupled to three independent 
quantum white noise fields $\xi_{in,j1}$, $\xi_{in,j2}$, 
and $\xi_{in,j3}$; 
they are continuous-mode and satisfy 
$[\xi_{in,jl}(t),\xi_{in,jl}(s)^*]=\delta(t-s)$. 
The oscillator modes satisfy the usual commutation relations 
$[a_j,a_k^*]=\delta_{jk}$, $[b_j,b^*_k]=\delta_{jk}$, $[a_j,b_k]=0$, 
and $[a_j,b_k^*]=0$ for $j,k=1,2$. 

In system $G_1$, the modes $a_1$ and $b_1$ are coupled via the 
two-mode squeezing Hamiltonian 
$H_1 = i\epsilon(a_1^*b^*_1-a_1 b_1)/2\sqrt{2}$ with $\epsilon$ 
constant. 
Moreover, $a_1$ is coupled to $\xi_{in,11}$ and $\xi_{in,12}$ via 
the coupling operators $L_{11}=\sqrt{\gamma} a_1$ and 
$L_{12}=i\sqrt{\kappa/2}a_1$, respectively, for some 
coupling constants $\gamma$ and $\kappa$. 
Also, $b_1$ is coupled to $\xi_{in,13}$ via 
$L_{13}=\sqrt{\kappa_1}b_1$ for coupling coefficient $\kappa_1$. 
We allow the possibility of losses in the two-mode squeezing 
process, which is modeled by the interaction of $a_1$ with the 
additional quantum noise field $\xi_{loss,11}$ via 
$L_{14}=\sqrt{\chi/2}a_1$ with $\chi$ a coupling constant. 
Similarly, $b_1$ interacts with $\xi_{loss,12}$ via 
$L_{15}=\sqrt{\chi/2}b_1$. 
The system $G_2$ has a similar structure. 
The modes $a_2$ and $b_2$ are coupled via the system Hamiltonian 
$H_2 = -\epsilon(a_2^*b^*_2 +a_2 b_2)/2\sqrt{2}$ with the same 
$\epsilon>0$. 
Also $a_2$ is coupled to $\xi_{in,21}$ and $\xi_{in,22}$ via 
$L_{21}=\sqrt{\gamma}a_2$ and 
$L_{22}=\sqrt{\kappa/2}a_2$, while $b_2$ is coupled to $\xi_{in,23}$ 
via $L_{23}=\sqrt{\kappa_1}b_2$. 
As in $G_1$, possible losses in the two-mode squeezing process in 
$H_2$ is modeled by coupling $a_2$ and $b_2$ to additional noise 
fields $\xi_{loss,21}$ and $\xi_{loss,22}$, respectively, via 
$L_{24}=\sqrt{\chi/2}a_2$ and $L_{25}=\sqrt{\chi/2}b_2$ with the 
same constant $\chi$ as before. 
The input fields $\xi_{in,j1}$, $\xi_{in,j3}$, $\xi_{loss,j1}$, 
and $\xi_{loss,j2}$ for $j=1,2$ are in the vacuum state. 

We allow the possibility of photon losses in the two transmission 
channels connecting $G_1$ and $G_2$. 
The losses are modeled by inserting in the two transmission 
paths a beam splitter with transmissivity $\alpha$ and reflectivity 
$\beta$, with $\alpha^2+\beta^2=1$. 
Each beam splitter BS$j$ has two ports, one for the 
incoming signal $\xi_{out,j3}$ and an unused port for a noise 
field $\xi_{BS,j}$. 
Moreover, we also allow the possibility of time delays along 
these transmission channels that are represented in the figure 
by the $\sigma_T$ blocks, with $T$ a positive number indicating 
the transmission delay. 
The time delay block acts on a signal $\xi$ coming into the 
block as $(\sigma_T \xi)(t)= \xi(t-T)$. 
Thus the interconnecting fields satisfy 
$\alpha(\sigma_T \xi_{out,13})(t)+\beta \xi_{BS,1}(t)=\xi_{in,22}(t)$ 
and $\alpha(\sigma_T \xi_{out,23})(t)+\beta \xi_{BS,2}(t)=\xi_{in,12}(t)$. 
The dynamics of the whole network is then given as follows:
\begin{align*}
  \dot{a}_1(t) 
    &=-(\frac{\gamma}{2}+\frac{\kappa}{4}+\frac{\chi}{4})a_1(t) 
       +\frac{\epsilon}{2\sqrt{2}}b_1(t)^* -\sqrt{\gamma}\xi_{in,11}(t) \\
  & \quad
       +i\sqrt{\frac{\kappa}{2}}
                 \Big[\alpha \Big\{
                    \sqrt{\kappa_1}(\sigma_T b_2)(t) 
                        + (\sigma_T \xi_{in,23})(t)\Big\} 
                            \\
  & \quad
       + \beta \xi_{BS,2}(t)\Big] -\sqrt{\frac{\chi}{2}}\xi_{loss,11}(t), \\
  \dot{b}_1(t) 
    & = \frac{\epsilon}{2\sqrt{2}}a_1(t)^*
        -(\frac{\kappa_1}{2}+\frac{\chi}{4})b_1(t)
           -\sqrt{\kappa_1}\xi_{in,13}(t) \\
  & \quad
        -\sqrt{\frac{\chi}{2}}\xi_{loss,12}(t), 
\end{align*}
\begin{align*}
    \dot{a}_2(t) 
      &=-(\frac{\gamma}{2}+\frac{\kappa}{4}+\frac{\chi}{4})a_2(t) 
          +\frac{i\epsilon}{2\sqrt{2}}b_2(t)^* 
             -\sqrt{\gamma}\xi_{in,21}(t) \\
   & \quad
          -\sqrt{\frac{\kappa}{2}}\Big[\alpha \Big\{
              \sqrt{\kappa_1}(\sigma_T b_1)(t) 
                 + (\sigma_T \xi_{in,13})(t) \Big\} \\
     &\quad + \beta \xi_{BS,1}(t)\Big]-\sqrt{\frac{\chi}{2}}\xi_{loss,21}(t), \\
        \dot{b}_2(t) 
      &= \frac{i\epsilon}{2\sqrt{2}}a_2(t)^*
         -(\frac{\kappa_1}{2}+\frac{\chi}{4})b_2(t)
            -\sqrt{\kappa_1}\xi_{in,23}(t) \\
      &\quad -\sqrt{\frac{\chi}{2}}\xi_{loss,22}(t), 
\end{align*}
with outputs
\begin{align*}
  \xi_{out,11}(t) 
     &= \sqrt{\gamma} a_1(t) + \xi_{in,11}(t),\\
   \xi_{out,12}(t) 
     &= i\sqrt{\frac{\kappa}{2}} a_1(t) 
            + \alpha\sqrt{\kappa_1}(\sigma_T b_2)(t) 
                 + \alpha (\sigma_T \xi_{in,23})(t) \\
  & \quad +\beta \xi_{BS,2}(t), 
\end{align*}
\begin{align*}
    \xi_{out,21}(t) 
      &= \sqrt{\gamma} a_2(t) + \xi_{in,21}(t), \\
   \xi_{out,22}(t) 
      & = \sqrt{\frac{\kappa}{2}} a_2(t) 
           +\alpha\sqrt{\kappa_1}(\sigma_T b_1)(t) 
              + \alpha (\sigma_T \xi_{in,13})(t) \\
   & \quad + \beta \xi_{BS,1}(t).
\end{align*}

\begin{figure*}[tbph]
\centering
\includegraphics[scale=0.65]{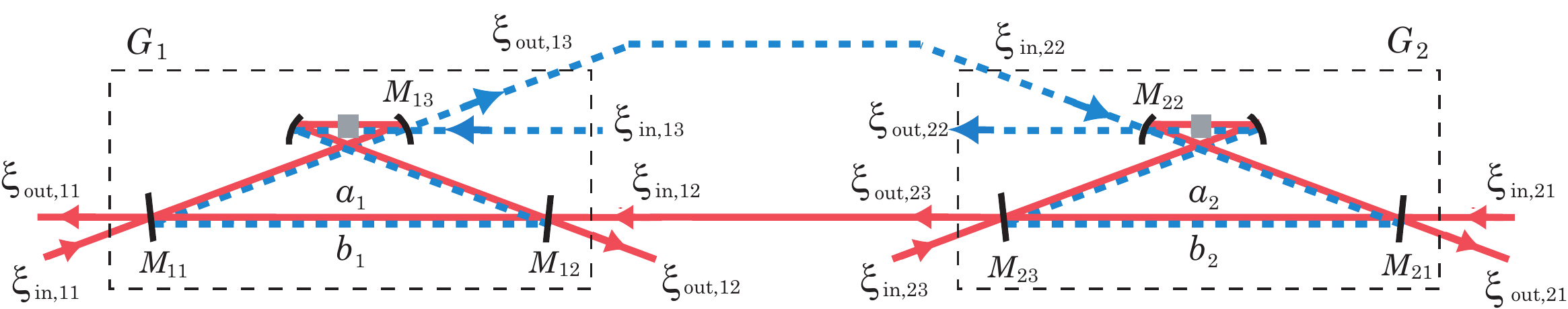}
\caption{
Candidate for a realization of the systems $G_1$ and $G_2$ 
in quantum optics. 
The modes $a_1$ and $b_1$ are frequency degenerate but orthogonally 
polarized. 
The mirrors composing the cavity $G_1$ are partially transmissive, 
depending on polarization of the light fields; the mirrors $M_{11}$ 
and $M_{12}$ are partially transmissive for $a_1$ but perfectly 
reflective for $b_1$, while $M_{13}$ is partially transmissive for 
$b_1$ but perfectly reflective for $a_1$. 
The cavity $G_2$ is also constructed in a similar way. 
For simplicity, other optical devices such as phase shifters are 
not shown. }
\label{fig:optics_realization}
\end{figure*}

The oscillator systems $G_1$ and $G_2$ could, in principle, be 
physically realized using optical cavities with $\chi^{(2)}$ nonlinear 
crystals, as depicted in Fig.~\ref{fig:optics_realization}. 
More precisely, the bow-tie type cavity $G_1$ contains two modes 
$a_1$ and $b_1$ that overlap and interact in a $\chi^{(2)}$ nonlinear 
crystal driven by a classical pump beam of effective amplitude 
$\epsilon/\sqrt{2}$. 
The modes $a_1$ and $b_1$ are frequency degenerate but orthogonally 
polarized. 
The mirrors composing the cavity are partially transmissive, 
depending on polarization of light fields; 
in particular, the mirrors $M_{11}$ and $M_{12}$ are partially 
transmissive for $a_1$ but perfectly reflective for $b_1$, 
while $M_{13}$ is partially transmissive for $b_1$ but perfectly 
reflective for $a_1$. 
The transmittance of $M_{11}$, $M_{12}$, and $M_{13}$ are 
$T_{11}=\gamma l/c$, $T_{12}=\kappa l/2c$, and $T_{13}=\kappa_1 l/c$, 
respectively, with $l$ the optical path length of the cavity and $c$ 
the speed of light. 
In addition, the field $\xi_{in,12}$ entering through mirror 
$M_{12}$ and the output $\xi_{out,12}$ are both passed through 
a 180$^{\rm o}$ phase shifter. 
The system $G_2$ is similarly realized by a bow-tie type cavity. 
For a more detailed description, see e.g., \cite{Iida2012}. 

The particular model described above is of interest because, in the 
large limit of the parameters that adiabatically eliminate  $b_1$ and 
$b_2$  \cite{Nurdin2009}, and taking the limit of zero time delays 
along the transmission channels \cite{Gough2009a,Gough2009b}, 
the remaining modes $a_1$ and $a_2$ couple via the Hamiltonian 
$\hat{H}= -i\alpha \kappa (a_1^*a_2^*-a_1a_2)$; 
this is a two-mode squeezing Hamiltonian  which is
the basis for generating entangled photon pairs in a nondegenerate 
optical parametric amplifier (NOPA) \cite{Ou1992}. 
In this sense, our system is a realistic approximation to the ideal 
system where two spatially separated systems effectively
interact through this two-mode squeezing Hamiltonian.


\subsection{Quadrature form and transfer function}
\label{sec:quad-uncontrolled}

We now specialize to the case where the transmission delay $T$ is 
negligible compared to the time scale of the dynamics of the 
systems $G_1$ and $G_2$. 
Then, in terms of quadratures, the coupled network Langevin equations 
with no transmission delays are:
\begin{align*}
    \dot{a}^q_{1} 
      &=-(\frac{\gamma}{2}+\frac{\kappa}{4}+\frac{\chi}{4})a^q_{1}
         +\frac{\epsilon}{2\sqrt{2}}b^q_{1} 
           -\alpha\sqrt{\frac{\kappa \kappa_1}{2}} b^p_{2} 
             -\sqrt{\gamma}\xi^q_{in,11}          \notag \\
      &\quad -\alpha \sqrt{\frac{\kappa}{2}}\xi^p_{in,23} 
      - \beta \sqrt{\frac{\kappa}{2}} \xi^p_{BS,2}
         -\sqrt{\frac{\chi}{2}} \xi^q_{loss,11}, \\
    \dot{a}^p_{1} 
      &=-(\frac{\gamma}{2}+\frac{\kappa}{4}+\frac{\chi}{4})a^p_{1}
         -\frac{\epsilon}{2\sqrt{2}}b^p_{1}
            +\alpha\sqrt{\frac{\kappa \kappa_1}{2}} b^q_{2} 
              -\sqrt{\gamma}\xi^p_{in,11}         \notag\\
      &\quad 
         +\alpha \sqrt{\frac{\kappa}{2}}\xi^q_{in,23} 
           +\beta \sqrt{\frac{\kappa}{2}} \xi^q_{BS,2}
              -\sqrt{\frac{\chi}{2}} \xi^p_{loss,11}, \\
    \dot{b}^q_1 
      &= \frac{\epsilon}{2\sqrt{2}}a^q_1
           -(\frac{\kappa_1}{2}+\frac{\chi}{4})b^q_1
              -\sqrt{\kappa_1}\xi^q_{in,13} 
                 -\sqrt{\frac{\chi}{2}} \xi^q_{loss,12}, \\
    \dot{b}^p_1 
      &= -\frac{\epsilon}{2\sqrt{2}}a^p_1
           -(\frac{\kappa_1}{2}+\frac{\chi}{4})b^p_1
              -\sqrt{\kappa_1}\xi^p_{in,13}  
                 -\sqrt{\frac{\chi}{2}} \xi^p_{loss,12}, 
\end{align*}
\begin{align*}
    \dot{a}^q_2 
      &=-(\frac{\gamma}{2}+\frac{\kappa}{4}+\frac{\chi}{4})a^q_2
          +\frac{\epsilon}{2\sqrt{2}}b^p_2 
            -\alpha\sqrt{\frac{\kappa \kappa_1}{2}} b^q_1 
               -\sqrt{\gamma}\xi^q_{in,21}             \notag \\
    &\quad -\alpha \sqrt{\frac{\kappa}{2}}\xi^q_{in,13} 
             + \beta \sqrt{\frac{\kappa}{2}} \xi^q_{BS,1} 
                -\sqrt{\frac{\chi}{2}} \xi^q_{loss,21}, \\
    \dot{a}^p_2 
      &=-(\frac{\gamma}{2}+\frac{\kappa}{4}+\frac{\chi}{4})a^p_2 
          +\frac{\epsilon}{2\sqrt{2}}b^q_2 
            -\alpha\sqrt{\frac{\kappa \kappa_1}{2}} b^p_1 
              -\sqrt{\gamma}\xi^p_{in,21}            \notag \\
      &\quad 
          -\alpha \sqrt{\frac{\kappa}{2}}\xi^p_{in,13} 
            -\beta \sqrt{\frac{\kappa}{2}} \xi^p_{BS,1} 
              -\sqrt{\frac{\chi}{2}} \xi^p_{loss,21}, 
 \end{align*}
 \begin{align*}             
    \dot{b}^q_2 
      &= \frac{\epsilon}{2\sqrt{2}}a^p_2
          -(\frac{\kappa_1}{2}+\frac{\chi}{4})b^q_2
            -\sqrt{\kappa_1}\xi^q_{in,23} 
              -\sqrt{\frac{\chi}{2}} \xi^q_{loss,22}, \\
    \dot{b}^p_2 
      &= \frac{\epsilon}{2\sqrt{2}}a^q_2
          -(\frac{\kappa_1}{2}+\frac{\chi}{4})b^p_2
            -\sqrt{\kappa_1}\xi^p_{in,23} 
              -\sqrt{\frac{\chi}{2}} \xi^p_{loss,22},
\end{align*}
with outputs 
\begin{align*}
    \xi^q_{out,11} &= \sqrt{\gamma} a^q_1 + \xi^q_{in,11}, ~~~
    \xi^p_{out,11} = \sqrt{\gamma} a^p_1 + \xi^p_{in,11}, \\
    \xi^q_{out,12} &= -\sqrt{\frac{\kappa}{2}} a^p_1 
       + \alpha\sqrt{\kappa_1}b_2^q +\alpha \xi^q_{in,23} 
         +\beta \xi^q_{BS,2}, \\
    \xi^p_{out,12} &= \sqrt{\frac{\kappa}{2}} a^q_1 
       + \alpha \sqrt{\kappa_1}b_2^p+\alpha \xi^p_{in,23} 
         + \beta \xi^p_{BS,2}, 
\end{align*}
\begin{align*}
   \xi^q_{out,21} &= \sqrt{\gamma} a^q_2 + \xi^q_{in,21}, ~~~
   \xi^p_{out,21} = \sqrt{\gamma} a^p_2 + \xi^p_{in,21}, \\
   \xi^q_{out,22} &= \sqrt{\frac{\kappa}{2}} a^q_2 
      + \alpha\sqrt{\kappa_1}b^q_1 + \alpha \xi^q_{in,13} 
        +\beta \xi^q_{BS,1}, \\
   \xi^p_{out,22} &= \sqrt{\frac{\kappa}{2}} a^p_2 
      + \alpha\sqrt{\kappa_1}b^p_1 
        + \alpha \xi^p_{in,13} 
          + \beta \xi^p_{BS,1}. 
\end{align*}
Now, let us observe some properties of the above dynamical 
equations for the oscillator and field quadratures. 
In particular, from the above equations it can be verified 
that the equations are not fully coupled and in fact:
(a) The set of equations for $a_1^q, a_2^q, b_1^q$, and $b_2^p$ 
form a closed set of equations driven by the {\em commuting} 
set of noises $\xi^q_{in,11}$, $\xi^p_{in,23}$, $\xi^q_{in,13}$, 
$\xi^q_{in,21}$, $\xi^q_{loss,11}$, 
$\xi^q_{loss,12}$, $\xi^q_{loss,21}$, $\xi^p_{loss,22}$, 
$\xi^q_{BS,1}$, and $\xi^p_{BS,2}$, and (b) the set of equations 
for $a_1^p, a_2^p, b_1^p$, and $b_2^q$ 
form another closed set of equations driven by the {\em commuting} 
set of noises $\xi^p_{in,11}$, $\xi^q_{in,23}$, $\xi^p_{in,13}$, 
$\xi^p_{in,21}$, $\xi^p_{loss,11}$, $\xi^p_{loss,12}$, 
$\xi^p_{loss,21}$, $\xi^q_{loss,22}$, $\xi^p_{BS,1}$, and $\xi^q_{BS,2}$. 
Although we are considering the case of no time delays, 
it may be easily inspected that the decoupling between the above 
sets of closed  equations for certain quadratures and commuting 
noises also holds in the time delay case since the structure 
of the equations are precisely the same. 

We can now consider the Fourier transforms of the observables 
above and the system transfer functions. 
Introduce the column vector of system operators 
\begin{align*}
z_1(t) &=(a_1^q(t),a_2^q(t),b_1^q(t),b_2^p(t), )^T,\\
z_2(t) &=(a_1^p(t), a_2^p(t),b_1^p(t),b_2^q(t))^T,\\
z(t) &=(z_1(t)^T,z_2(t)^T)^T,
\end{align*}
and 
\begin{align*}
   \xi_1
     &=(\xi^q_{in,11}, \xi^p_{in,23}, \xi^q_{in,13}, 
         \xi^q_{in,21}, \\
   &\quad  \xi^q_{loss,11}, \xi^q_{loss,12}, \xi^q_{loss,21}, 
         \xi^p_{loss,22}, \xi^q_{BS,1},\xi^p_{BS,2})^T,\\
   \xi_2
     &=(\xi^p_{in,11},\xi^q_{in,23},\xi^p_{in,13}, 
         \xi^p_{in,21},\\
   &\quad \xi^p_{loss,11}, \xi^p_{loss,12}, \xi^p_{loss,21}, 
         \xi^q_{loss,22}, \xi^p_{BS,1},\xi^q_{BS,2})^T,\\
   \xi &= (\xi_1(t)^T,\xi_2(t)^T)^T.
\end{align*}
By the decoupling structure already noted above, we have 
\begin{align*}
   & \dot{z}_j(t) = A_j z_j(t) + B_j\xi_j(t),~~~(j=1,2), \\
   & \xi^q_{out,11}(t)+\xi^q_{out,21}(t) = C_1 z_1(t) + D_1\xi_1(t), \\
   & \xi^p_{out,11}(t)-\xi^p_{out,21}(t) = C_2 z_2(t) + D_2\xi_2(t),
\end{align*}
where $A_1, A_2, B_1$, and $B_2$ are real matrices of suitable 
dimensions whose entries can be readily determined from the 
equations for $a_j^q, a_j^p, b_j^q$, and $b_j^p$, and the 
row vectors $C_1$, $C_2$, $D_1$, and $D_2$ are given by:
\begin{align*}
   C_1 &= (1,~ 1,~ 0_{1 \times 2}), ~~~
   C_2 = (1,~ -1,~ 0_{1 \times 2}), \\
   D_1 &= (1,~ 0_{1 \times 3},~ 1,~ 0_{1\times 6}), ~~~
   D_2 = (1,~ 0_{1 \times 3},~ -1,~ 0_{1 \times 6}). 
\end{align*}

Let $x_j(t)=\xi^q_{out,j1}(t)$ and $y_j(t)=\xi^p_{out,j1}(t)$ and 
let $X_j$, $Y_j$, and $\Xi_j$ denote the Fourier transforms of 
$x_j$, $y_j$, and $\xi_j$, respectively (see Section 
\ref{sec:entanglement-criterion}). 
Then 
\begin{align*}
    & X_1(i\omega)+X_2(i\omega)
      =H_1(i\omega) \Xi_1(i\omega), \\
    & Y_1(i\omega)-Y_2(i\omega)
      =H_2(i\omega) \Xi_2(i\omega), 
\end{align*}
with $H_j(i\omega)=C_j(i\omega I-A_j)^{-1}B_j+D_j$ ($j=1,2$). 
Using the fact that 
$\langle (\Xi_j(i\omega)^*)^T \Xi_j(i\omega')^T \rangle 
= \langle \Xi_j(-i\omega)\Xi_j(i\omega')^T \rangle 
= I_{10 \times 10} \delta (\omega-\omega')$, 
we find that 
$\langle (X_1(i\omega)+X_2(i\omega))^*(X_1(i\omega')+X_2(i\omega')) \rangle 
= {\rm Tr}[H_1(i\omega)^*H_1(i\omega)]\delta(\omega-\omega')$ and 
$\langle (Y_1(i\omega)-Y_2(i\omega))^*(Y_1(i\omega')-Y_2(i\omega')) \rangle 
= {\rm Tr}[H_2(i\omega)^*H_2(i\omega)]\delta(\omega-\omega')$. 
Thus, for the uncontrolled network we conclude that 
\begin{align*}
   & V_+(i\omega)={\rm Tr}[H_1(i\omega)^*H_1(i\omega)],  \\
   & V_-(i\omega)={\rm Tr}[H_2(i\omega)^*H_2(i\omega)].
\end{align*}


\section{LQG feedback control of the system}

\begin{figure*}[tbph]
\centering
\includegraphics[scale=0.7]{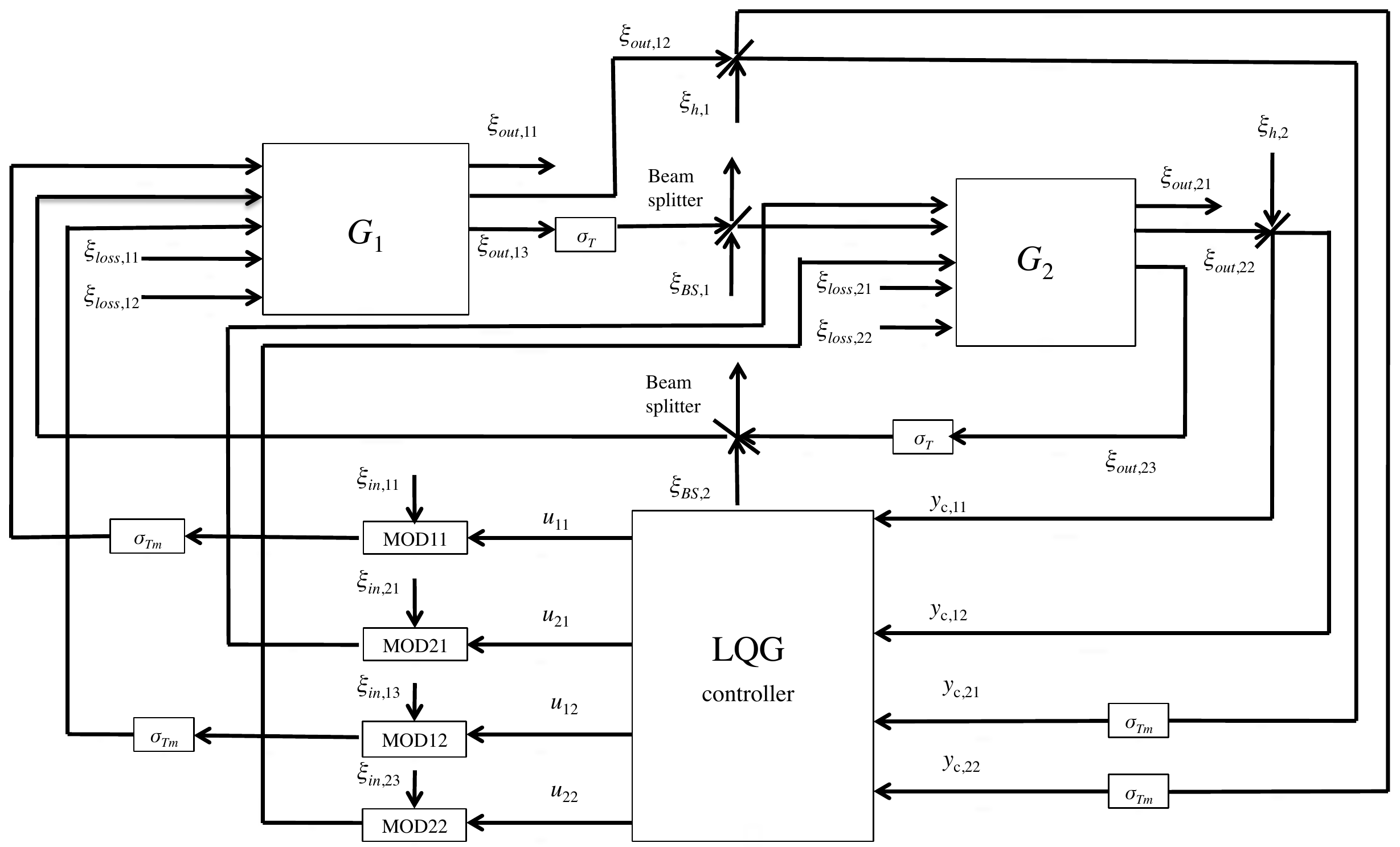}
\caption{
The controlled quantum network.
The LQG controller is located on the site of $G_2$; 
hence the time delays of communication between the controller 
and the system $G_2$ is negligible, while we assume that 
the control signals $u_{11}$ and $u_{12}$ propagate from the 
controller to $G_1$ with time delays $T_m$. 
The operation bringing this time-delay is denoted by $\sigma_{T_m}$. 
}
\label{fig:controlled_net}
\end{figure*}

The control scheme is shown in Fig.~\ref{fig:controlled_net}. 
The input to the controller will be the signals $y_{c,11}$, 
$y_{c,12}$, $y_{c,21}$, and $y_{c,22}$ that are obtained by 
performing dual homodyne detection on the output fields 
$\xi_{out,12}$ and $\xi_{out,22}$, respectively. 
Two dual homodyne detectors, labeled 1 and 2, are required, 
and each consists of a 50:50 beam splitter, where at the unused 
beam splitter port a vacuum noise source $\xi_{h,j}, (j=1,2)$ 
comes in. 
At the output of one beam splitter, the position quadrature of 
the field is measured while at the other output the momentum 
quadrature is measured. 
For the $j$-th dual homodyne detector, the outputs $y_{c,j1}$ 
and $y_{c,j2}$ are given by:
\begin{align}
    y_{c,11}(t) 
    &=\frac{\sqrt{\kappa}}{2}a^q_2(t)
        +\alpha\sqrt{\frac{\kappa_1}{2}}b^q_1(t)
           +\frac{\alpha}{\sqrt{2}} \xi^q_{in,13}(t) \notag \\
    &\quad  +\frac{\beta}{\sqrt{2}}\xi^q_{BS,1}(t)
        +\frac{1}{\sqrt{2}}\xi^q_{h,2}(t), \label{eq:heterodyne-start} 
\\
    y_{c,12}(t) 
      &=-\frac{\sqrt{\kappa}}{2}a^p_2(t)
        -\alpha\sqrt{\frac{\kappa_1}{2}}b^p_1(t)
           -\frac{\alpha}{\sqrt{2}} \xi^p_{in,13}(t) \notag \\
      &\quad  -\frac{\beta}{\sqrt{2}}\xi^p_{BS,1}(t)
       +\frac{1}{\sqrt{2}}\xi^p_{h,2}(t), \\
    y_{c,21}(t) 
      &=-\frac{\sqrt{\kappa}}{2}(\sigma_T a^p_1)(t)
       +\alpha\sqrt{\frac{\kappa_1}{2}}(\sigma_T b^q_2)(t) \notag \\
      &\quad +\frac{\alpha}{\sqrt{2}} (\sigma_T \xi^q_{in,23})(t) 
       +\frac{\beta}{\sqrt{2}}(\sigma_T \xi^q_{BS,2})(t) \notag\\
      &\quad  +\frac{1}{\sqrt{2}}(\sigma_T \xi^q_{h,1})(t), 
\end{align}
\begin{align}
    y_{c,22}(t) 
      &=-\frac{\sqrt{\kappa}}{2}(\sigma_T a^q_1)(t)
         -\alpha\sqrt{\frac{\kappa_1}{2}}(\sigma_T b^p_2)(t)\notag \\
      &\quad  -\frac{\alpha}{\sqrt{2}} (\sigma_T \xi^p_{in,23})(t)
         -\frac{\beta}{\sqrt{2}}(\sigma_T \xi^p_{BS,2})(t) \notag\\
      &\quad +\frac{1}{\sqrt{2}}(\sigma_T \xi^p_{h,1})(t),
\label{eq:heterodyne-end}
\end{align}
The controller produces 8 control signals $u^q_{j1}(t)$, $u^p_{j1}(t)$, 
$u^q_{j2}(t)$, and $u^p_{j2}(t)$ for $j=1,2$ which drive the 
quantum system via modulators. 
The control signals $u^q_{jl}$ and $u^p_{jl}$ form the real and 
imaginary component of the complex classical signal 
$u_{jl}=u_{jl}^q+i u^p_{jl}$ that will drive the modulator 
labeled MOD$jl$ for $j,l=1,2$. 
The output of the modulator MOD$jl$ then drives the system $G_j$, 
see Fig.~\ref{fig:controlled_net}. 
The position and momentum quadratures of the output field of MOD$j1$ 
are $u^q_{j1}(t) + \xi^q_{in,j1}$ and $u^p_{j1}(t) + \xi^p_{in,j1}$, 
respectively, while the position and momentum quadratures of the 
output field MOD$j2$ are $u^q_{j2}(t) + \xi^q_{in,j3}$ and 
$u^p_{j2}(t) + \xi^p_{in,j3}$, respectively. 
Let $y_c(t)=(y_{c,11}(t),y_{c,12}(t),y_{c,21}(t),y_{c,22}(t))^T$ and 
\begin{align*}
   u_c = (u^q_{11}, u^p_{11}, u^q_{21}, u^p_{21}, 
          u^q_{12}, u^p_{12}, u^q_{22}, u^p_{22})^T.
\end{align*}
The LQG controller has an internal 8-th order state $z_c(t)$ that 
obeys the following classical Langevin equation:
\begin{align}
\label{eq:controller-internal}
    \dot{z}_c(t) = A_c z_c(t) + B_c y_c(t), ~~~
    u_c(t) = C_c z_c(t), 
\end{align}
where $A_c$, $B_c$, $C_c$ are real matrices of the appropriate 
dimensions. 
Here, the order of $z_c(t)$ is 8, since the degree of the system 
to be controlled is 8, corresponding to the oscillator quadratures 
$a_j^q$, $a_j^p$, $b_j^q$, and $b_j^p$ for $j=1,2$. 

We assume that the controller is located on the site of $G_2$ so 
that the delays in transmitting the control signals $u_{21}$ and 
$u_{22}$ from the controller to the system $G_2$ are negligible. 
However, we allow the possibility of delays in transmitting the 
control signals $u_{11}$ and $u_{12}$ from the controller to $G_1$; 
here we assume those time delays take the same quantities and 
let us denote them by $T_m$. 
Then, the dynamical equation for the closed loop system can be 
obtained simply by making the substitutions of Eqs. 
(\ref{eq:heterodyne-start})-(\ref{eq:heterodyne-end}) into Eq. 
(\ref{eq:controller-internal}) and the substitutions 
\begin{align*}
   & \sigma_{T_m}(u_{11}+\xi_{in,11}) \rightarrow \xi_{in,11}, ~~
     \sigma_{T_m}(u_{12}+\xi_{in,13}) \rightarrow \xi_{in,13}, \\
   & u_{21}+\xi_{in,21} \rightarrow \xi_{in,21}, ~~
     u_{22}+\xi_{in,23} \rightarrow \xi_{in,23}
\end{align*}
into the dynamical equations of the system. 
Now define 
\[
    \tilde z=(z^T,z_c^T)^T, ~~~
    \tilde \xi=(\xi^T,\xi^q_{h,1},\xi^p_{h,1},
                       \xi^q_{h,2},\xi^p_{h,2})^T.
\]
The closed-loop system without time delays is then given by 
\begin{align*}
    \dot{\tilde z}(t) = \tilde A \tilde z(t) + \tilde B \tilde \xi(t),
\end{align*}
where 
$\tilde A$ and $\tilde B$ are real matrices of the form: 
\[
     \tilde A 
       = \left(\begin{array}{ccc} 
            A_1 & 0 & A_{13} \\ 
            0 & A_2 & A_{23} \\
            A_{31} & A_{32} & A_{33}
         \end{array} \right),~~
     \tilde B 
       = \left( \begin{array}{ccc} 
            B_1 & 0 & 0 \\
            0 & B_2 & 0\\ 
            B_{31} & B_{32} & B_{33}\\
         \end{array} \right).
\]
Here, $A_1$, $A_2$, $B_1$, and $B_2$ are as defined previously 
in Sec. \ref{sec:quad-uncontrolled}, and $A_{jl}$ and $B_{jl}$ 
are matrices that are determined by the resulting closed-loop 
system. 
The equations for $\xi^q_{out,11}+\xi^q_{out,21}$ and 
$\xi^p_{out,11}-\xi^p_{out,21}$ are now of the form: 
\begin{align*} 
    \xi^q_{out,11}(t)+\xi^q_{out,21}(t) 
       & = (C_1,~0_{1 \times 4},~C_{12})z(t) + \tilde D_1\tilde \xi(t), \\
    \xi^p_{out,11}(t)-\xi^p_{out,21}(t) 
       & = (0_{1 \times 4},~C_2,~C_{22})z(t)+ \tilde D_2\tilde \xi(t),
\end{align*}
with $C_1$ and $C_2$ also as given in 
Sec. \ref{sec:quad-uncontrolled}, and 
$\tilde D_1=(D_1,~0_{1 \times 14})$ and 
$\tilde D_2=(0_{1 \times 10},~D_2,~0_{1 \times 4})$. 
Note that there are contributions from the classical 
controller state $z_c(t)$ of the controller in these quadratures. 
However, since we are interested in the entanglement between 
the output fields, we can omit these contributions, since they, 
being classical, have no bearing on the degree of entanglement. 
Thus, in the case of a measurement-feedback controller in the 
loop it suffices to consider the modified outputs
\begin{align*} 
    \tilde \xi^q_{out,11}(t)+ \tilde \xi^q_{out,21}(t) 
       & = (C_1,~0_{1 \times 12})z(t) + \tilde D_1 \tilde \xi(t), \\
    \tilde \xi^p_{out,11}(t)-\tilde \xi^p_{out,21}(t) 
       & = (0_{1 \times 4},~C_2,~0_{1 \times 8})z(t) 
              + \tilde D_2 \tilde \xi(t).
\end{align*}
Define $x_1(t)=\tilde \xi^q_{out,11}(t)$, 
$x_2(t)=\tilde \xi^q_{out,21}(t)$, $y_1(t)=\tilde \xi^p_{out,11}(t)$, 
and $y_2(t)=\tilde \xi^p_{out,21}(t)$ and let 
$X_1(i\omega)$, $X_2(i\omega)$, $Y_1(i\omega)$, and $Y_2(i\omega)$ 
be their Fourier transforms, respectively. 
Also, let $\tilde \Xi(i\omega)$ denote the Fourier transform of 
$\tilde \xi(t)$. 
Then, the transfer function $\tilde H_1(i\omega)$ from 
$\tilde \Xi(i\omega)$ to $X_1(i\omega)+X_2(i\omega)$ and 
$\tilde H_2(i\omega)$ from $\tilde \Xi(i\omega)$ to 
$Y_1(i\omega)-Y_2(i\omega)$ are given by 
\begin{align*}
    \tilde H_1(i\omega) 
      &= (C_1,~0_{1 \times 12})(i\omega I-\tilde A)^{-1}\tilde B 
             + \tilde D_1, \\
    \tilde H_2(i\omega) 
      &= (0_{1 \times 4},~C_2,~0_{1 \times 8})
            (i\omega I-\tilde A)^{-1}\tilde B + \tilde D_2.
\end{align*}
Then, noting that $\langle  (\tilde \Xi(i\omega)^*)^T \tilde \Xi(i\omega')^T \rangle 
= \langle \tilde \Xi(-i\omega) \tilde \Xi(i\omega')^T \rangle = (I_{24 \times 24}+iZ) \delta(\omega-\omega')$, we obtain the 
expressions:
\begin{align*}
   & \lefteqn{\langle (X_1(i\omega)+X_2(i\omega))^*
                (X_1(i\omega')+X_2(i\omega')) \rangle} \\
       &\quad ={\rm Tr}\bigl[ (\tilde H_1(i\omega)^*\tilde H_1(i\omega))^T (I_{24 \times 24}+iZ) \bigr] \delta(\omega-\omega'), \\
              &\quad = {\rm Tr}\big[ \tilde H_1(i\omega)^* \tilde H_1(i\omega) \big] \delta(\omega-\omega'),
\end{align*}
\begin{align*}              
   & \lefteqn{\langle (Y_1(i\omega)-Y_2(i\omega))^*
                (Y_1(i\omega')-Y_2(i\omega')) \rangle} \\
       &\quad ={\rm Tr}\bigl[ (\tilde H_2(i\omega)^*\tilde H_2(i\omega))^T (I_{24 \times 24}+iZ) \bigr] \delta(\omega-\omega'),\\
       &\quad ={\rm Tr}\big[ \tilde H_2(i\omega)^* \tilde H_2(i\omega) \big] \delta(\omega-\omega').
\end{align*}
Here, $Z$ is a real invertible $24 \times 24$ skew-symmetric matrix given by $Z = \frac{1}{2i}\langle \tilde \Xi(-i\omega) \tilde \Xi(i\omega)^T- (\tilde \Xi(-i\omega)\tilde \Xi(i\omega)^T )^T \rangle$. 
Therefore, we conclude that the expressions for $V_+(i\omega)$ and 
$V_-(i\omega)$ when there is an LQG controller in the loop are 
\begin{align*}
    V_+(i\omega) 
       &={\rm Tr}\big[ \tilde H_1(i\omega)^* \tilde H_1(i\omega)  \big], \\
    V_-(i\omega) 
       &={\rm Tr}\big[ \tilde H_2(i\omega)^* \tilde H_2(i\omega) \big]. \\
\end{align*}
%


\section{Entanglement between continuous-mode Gaussian output fields: 
The ideal case}
\label{sec:results-ideal}

We here consider the entanglement generated between $\xi_{out,11}(t)$ 
and $\xi_{out,21}(t)$, with and without an LQG feedback controller, 
in the idealized situation where there are no losses in the two mode 
squeezing processes (i.e., $\chi=0$) and there are also no losses 
in the transmission channels between $G_1$ and $G_2$ (i.e., $\alpha=1$). 
First, we assume that the transmission delays along the transmission 
channels are negligible. 
Then we will consider how the control performance and the closed-loop 
stability are affected by non-negligible time delays. 


\subsection{Negligible transmission delays}
\label{sec:ideal-no-delays}

\begin{figure*}[tbph]
\centering
\includegraphics[scale=0.45]{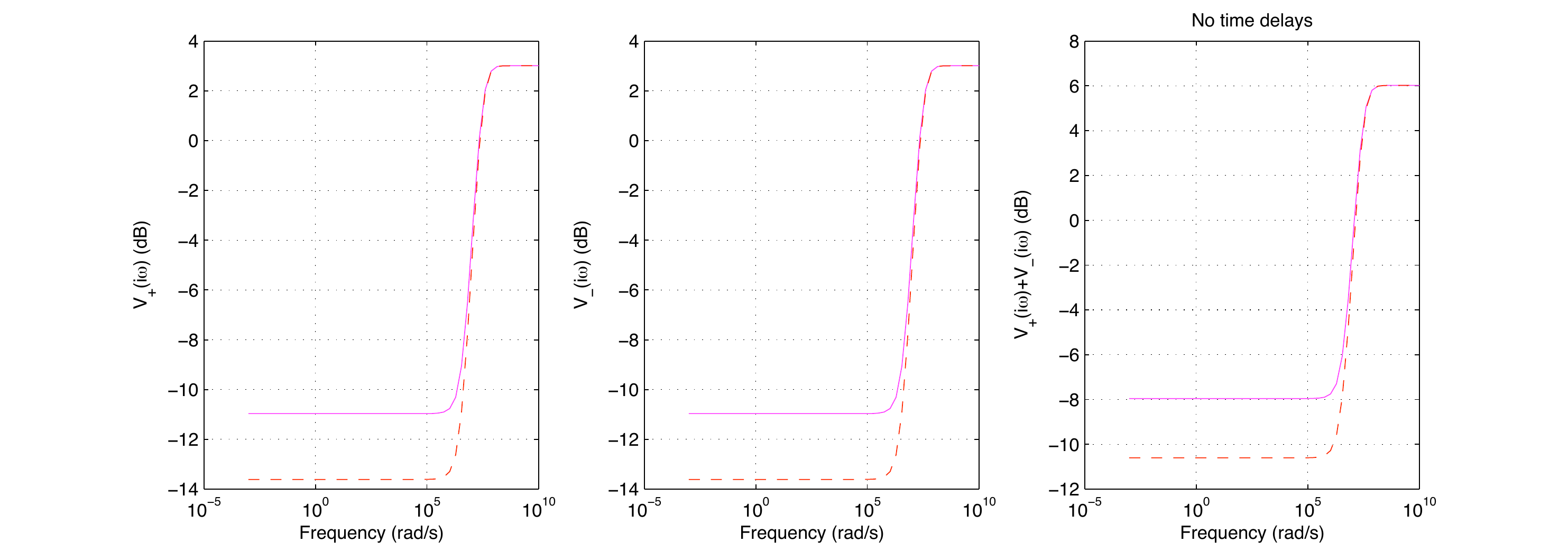}
\caption{Plots of $V_+(i\omega)$ (left), $V_-(i\omega)$ (middle), 
and $V_-(i\omega)+V_+(i\omega)$ (right) in dB against the 
frequency in rad/s, without an LQG controller (solid line) and 
with an LQG controller (dashed line) designed to minimize the 
cost function (\ref{eq:lqg-cost}). 
The  same lines are used as legends in subsequent figures as well. 
}
\label{fig:1}
\end{figure*}

Throughout we will consider the case where $\kappa = 1.8 \times 10^7$ Hz, 
$\gamma=1.5 \times \kappa = 2.7 \times 10^{7}$ Hz,  
$\kappa_1=10\times \kappa=1.8  \times 10^8$ Hz, 
and $\epsilon/\sqrt{2} = \sqrt{\kappa \kappa_1/2} 
=4.0249\times 10^7$ Hz. 
In the case of the optical system shown in Fig. 
\ref{fig:optics_realization}, these parameter values are realized 
using the mirrors with transmittance $T_{11}=T_{21}=0.045$, 
$T_{12}=T_{22}=0.015$, and $T_{13}=T_{23}=0.3$, where the optical 
path lengths of each cavities are both set to $l=0.5$ m. 
In this section we also set $\alpha=1$ and $T=T_m=0$. 
To design an 8th-order LQG controller, let us set the cost function 
\eqref{LQG cost} in the following form:
\begin{align}
    J(u_c) 
      &= \mathop{\lim}_{T \rightarrow \infty} \frac{1}{T}
            {\mathbb E}\Big[
                \int_{0}^T \Big\{ 
                    \varrho\big(\left[ \begin{array}{cc} C_1 & C_2\end{array} \right]z(t)\big)^2 \notag \\
      &\qquad + u_c(t)^Tu_c(t)\Big\} dt \Big]. 
\label{eq:lqg-cost}
\end{align}
The weighting constant $\varrho$ is taken to be $\varrho=1 \times 10^7$. 
With the use of the Matlab Control System Toolbox, we obtain 
the optimal LQG controller, 
and we show the frequency domain power spectra plots $V_+(i\omega)$, 
$V_-(i\omega)$, and $V_+(i\omega)+V_-(i\omega)$ in Fig.~\ref{fig:1} 
in dB scale (here $A \geq 0$ in linear scale is $10 \log_{10} A$ in dB scale). 
These figures indicate that all three power spectra are lower 
when the LQG controller is present as compared to when there 
is no controller. 
Here the controller can provide an additional attenuation 
to all three spectra by slightly more than 2.6 dB up to frequency 
of about $10^6$ rad/s. 
Moreover, by the entanglement criterion (\ref{eq:entangle-criterion}),
we see that entanglement is achieved for modes with frequency up to 
slightly above $10^7$ rad/s but less than $10^{8}$ rad/s (note that 
entanglement is achieved at a mode of frequency $\omega$ whenever 
$V_+(i\omega)+V_-(i\omega)<10\log_{10}(4)\,\hbox{dB} = 6.0206 \,\hbox{dB}$).
Note that this is achieved despite the controller being designed 
to minimize the cost function (\ref{eq:lqg-cost}) rather than 
minimizing any of the three power spectra directly, indicating 
the utility of the  cost function (\ref{eq:lqg-cost}) for feedback control of output entanglement. 


\subsection{Non-negligible transmission delays}
\label{sec:ideal-with-delays} 

\begin{figure*}[tbph]
\centering
\includegraphics[scale=0.45]{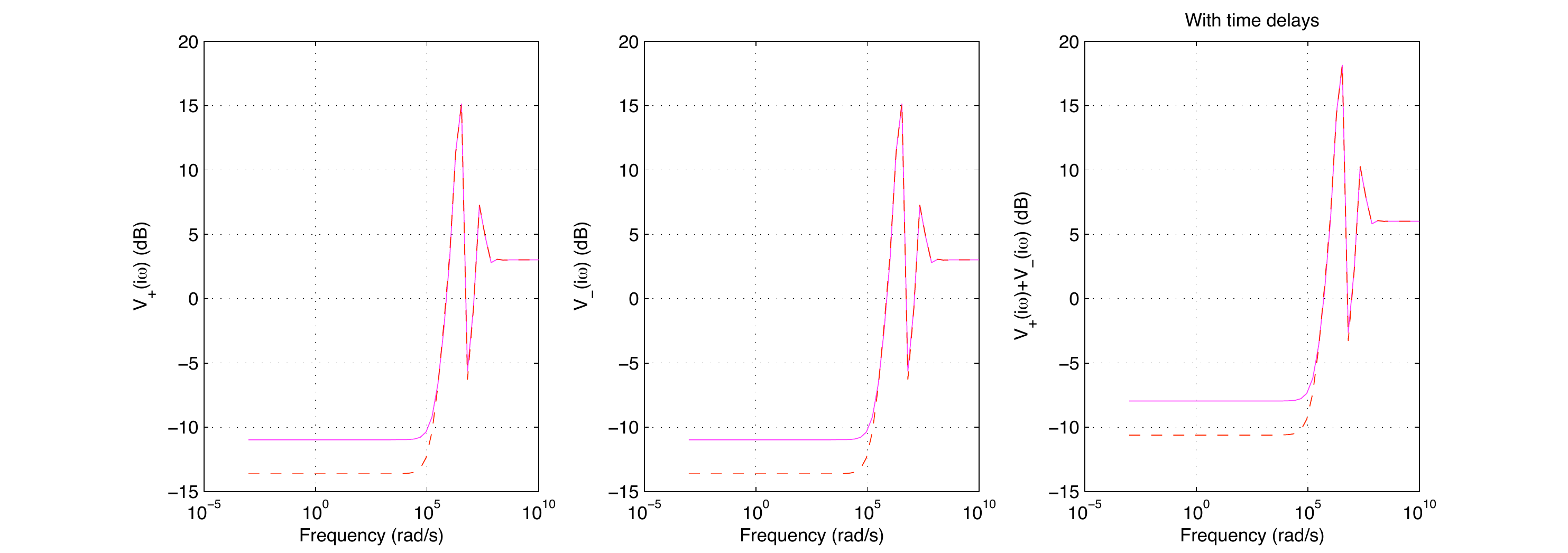}
\caption{Plots of $V_+(i\omega)$ (left), $V_-(i\omega)$ (middle), 
and $V_-(i\omega)+V_+(i\omega)$ (right) in dB against the 
frequency in rad/s when there are time delays present 
($T=1\times 10^{-6}$~s and $T_m=2 \times 10^{-6}$~s). 
}
\label{fig:2}
\end{figure*}

Since the dynamics of $z(t)$ and $z_c(t)$ preserves the Gaussian 
nature of the state, we can study an equivalent linear classical 
delayed differential Gaussian system with finite time delays (in the sense that the
mean and (symmetrized) covariance matrix of the quantum system and its classical equivalent evolve
in an identical manner). See, 
e.g., the text \cite{Michiels2007} for a treatment of classical linear systems with time delays. 
Such linear differential systems with delays can be handled 
with the Matlab Control System Toolbox's `delayss' object. 
We consider the case where $T = 10^{-6}$~s and 
$T_m=2 \times 10^{-6}$ s, which is about an order of magnitude 
longer than the time scale of the system dynamics. 
Using the controller that had been designed in Sec. 
\ref{sec:ideal-no-delays}, we obtain plots of $V_+(i\omega)$, 
$V_-(i\omega)$, and  $V_+(i\omega)+V_-(i\omega)$ as shown 
in Fig.~\ref{fig:2}. 
They indicate that despite the time delays there is still 
a significant reduction of about 2.6 dB in all three spectra 
for frequencies up to about $10^4$ rad/s and 1 dB for frequencies 
between $10^4$ and $10^5$ rad/s. 
Slightly above $10^5$ rad/s, there is no more reduction and in fact 
one can see a marked increase in the magnitude of each spectra 
at certain frequencies above $10^5$ rad/s. 
This suggests that the feedback controller is still effective 
for enhancing the entanglement between the output fields 
even in the presence of time delays, but the bandwidth at which 
enhancement is achieved is reduced. 
However, additional care has to be taken before we can be conclusive 
about this claim. 
As we have seen, there are frequency ranges in which all three 
spectra experience a large increase, which could be indicative of 
instability (instability here is in the sense that the symmetrized covariance
matrix of the system diverges as $t \rightarrow \infty$). 
Linear delay differential systems can be described by abstract 
infinite-dimensional differential equations \cite{Michiels2007}, 
and numerical algorithms 
are available to examine the stability of these systems. 
Here we use the freely available DDE-BIFTOOL toolbox 
\cite{Engelborghs2001,Engelborghs2002}, a Matlab toolbox to determine the stability 
of a delay differential system, and find that the system under consideration is indeed stable, 
the right hand most root of the characteristic equation 
of the system has a negative real part \footnote{Strictly speaking, DDE-BIFTOOL only checks
for the stability of the {autonomous} linear delay differential 
system that is obtained when all driving noises are set to 0. 
It is known that if all eigenvalues of the 
characteristic equation of the autonomous system have real parts less than some constant $-\alpha$ ($\alpha>0$) then its state decays to 0 exponentially fast, see, e.g., 
\cite[Lemma 1]{Cao2003}. However, this implies that the noise 
driven linear differential delay system is stable in the sense 
that the symmetrized covariance matrix converges as $t \rightarrow \infty$ by a straightforward and minor 
extension of the calculations presented in \cite[Section 3.2]{Lei2007}.}. 

Note that the log-log nature of the plots and the number of points used to produce Fig.~\ref{fig:2} (as well as Fig.~\ref{fig:20})  in Matlab give the impression  of non-smooth functions, but in fact this is {\em not} the case.  Since the analysis shows that the closed-loop system is stable,  the closed-loop transfer function has no poles on the right half plane and all poles are bounded away from the imaginary axis and thus the functions $V_{\pm}(i\omega)$ depicted in the plots are theoretically guaranteed to be smooth functions of $\omega$ (i.e., they infinitely differentiable functions of $\omega$). However, the plots indicate that when there are time delays these functions fluctuate faster at some high frequencies. The fluctuations in the figures are prominent since the time delays have been taken to be comparable to the time scale of the node dynamics. If the delays are gradually decreased to zero then the high frequency fluctuations gradually smooths out.


\section{Effect of amplification losses and transmission losses}
\label{sec:results-realistic}

Here we consider the more realistic case where there are losses in the 
two-mode squeezing process and  along the transmission lines 
connecting $G_1$ and $G_2$. 
We will design our LQG controller based on the assumption 
that we do not how much these losses appear in the system, 
effectively. Thus we set the controller to be identical to the one 
we had previously designed under the assumption $\chi=0$ and $\alpha=1$. 

\begin{figure}[tbph]
\centering
\includegraphics[scale=0.34]{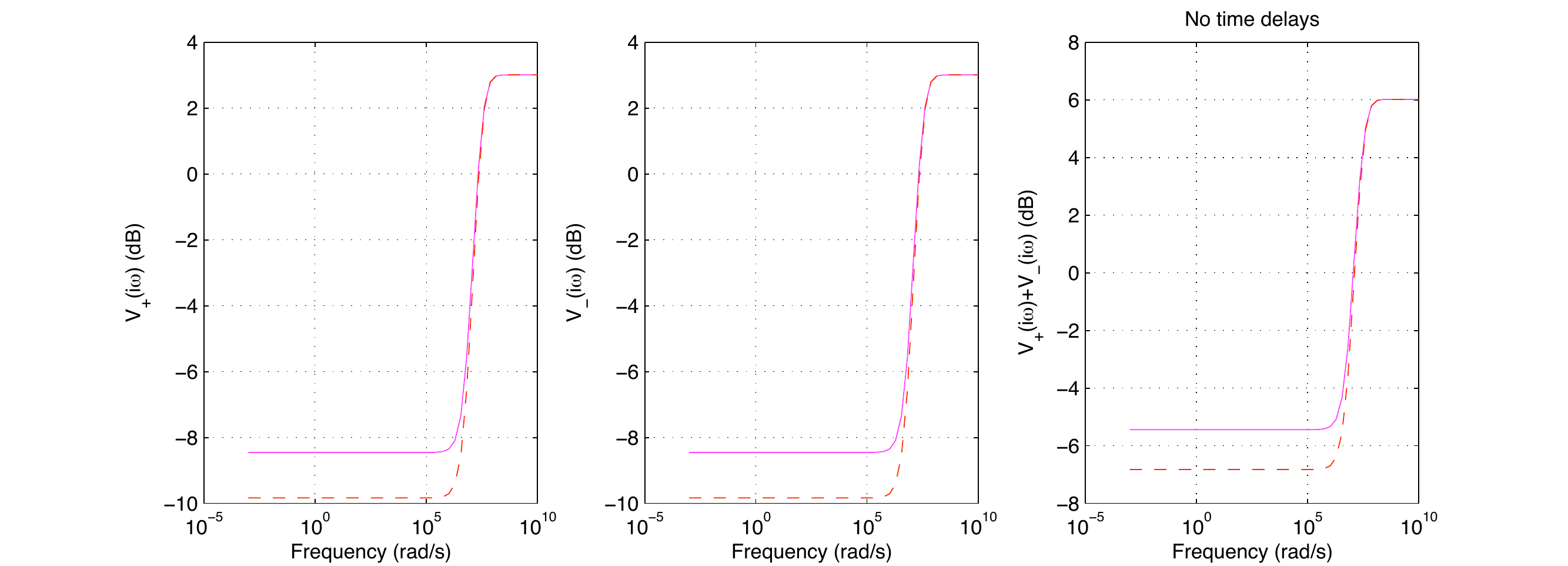}
\caption{Performance of the system when the amplification loss 
is $\chi=1.3975\times 10^6$ Hz, no transmission losses, 
and there are no delays. 
}
\label{fig:3}
\end{figure}

\begin{figure}[tbph]
\centering
\includegraphics[scale=0.35]{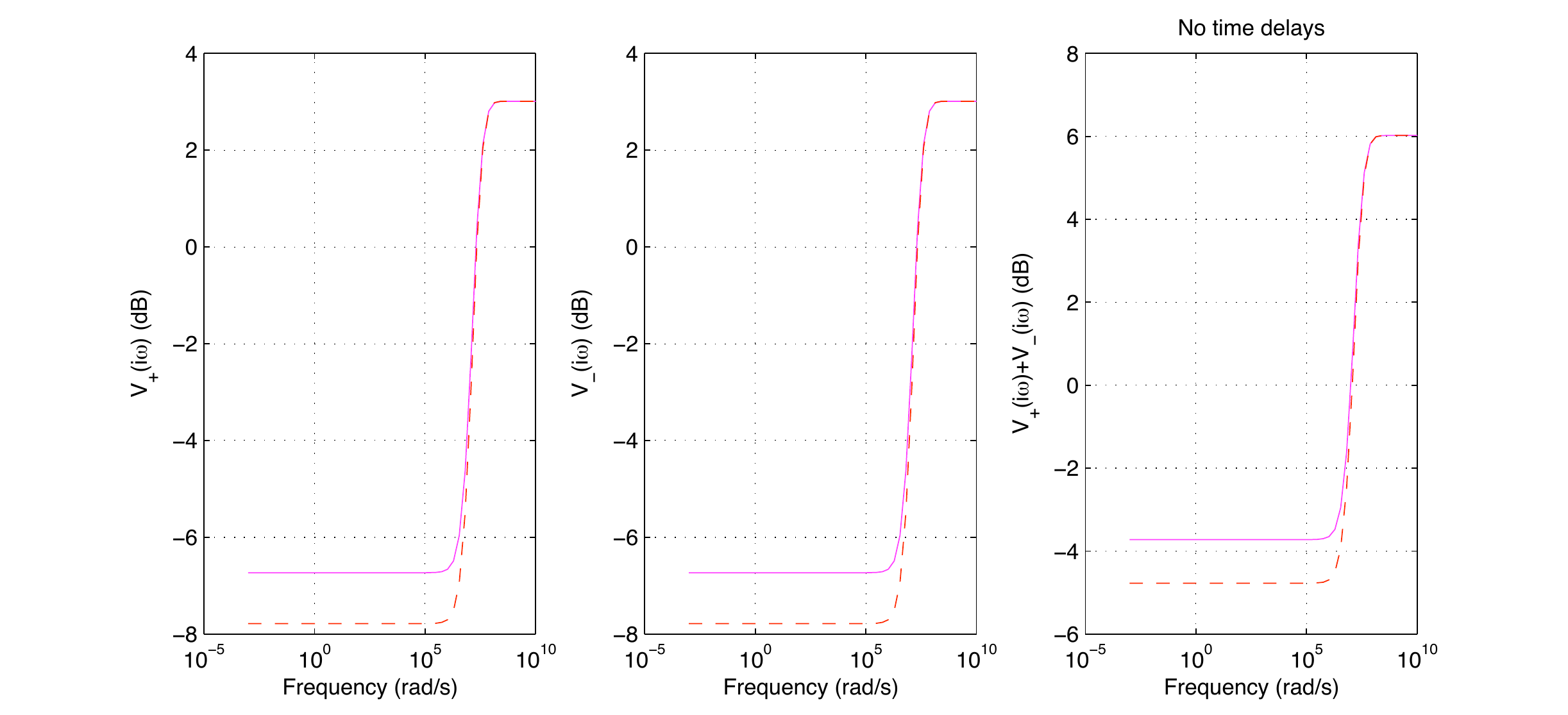}
\caption{Performance of the system  when the amplification loss 
coefficient is $\chi=1.3975\times 10^6$ Hz, transmission losses 
of 3\%, and there are no delays. 
}
\label{fig:5}
\end{figure}

\begin{figure}[tbph]
\centering
\includegraphics[scale=0.3]{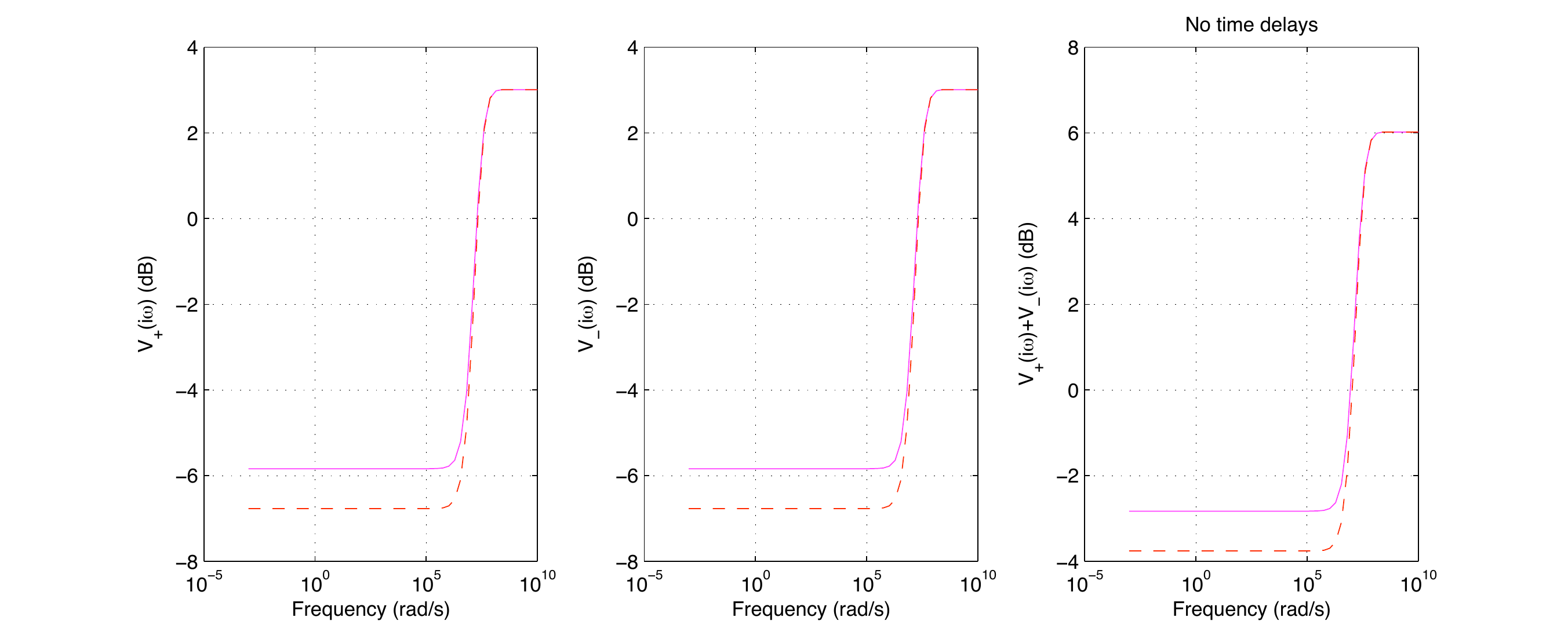}
\caption{Performance of the system when the amplification loss 
coefficient is $\chi=1.3975 \times 10^6$ Hz, transmission losses of 
5\%, and there are no delays. 
}
\label{fig:7}
\end{figure}

We consider first the case where the amplification loss 
coefficient is $\chi = 1.3975 \times 10^6$ Hz and there are no 
transmission losses ($\alpha=1$). 
The results are shown in Fig.~\ref{fig:3}. It can be seen that 
in the frequency region where the controller can reduce the power 
spectra, the reduction is smaller than if there were no amplification 
losses (about 1.4 dB reduction compared to about 2.6 dB reduction 
in the latter).
For the same amplification loss coefficient, the cases with 
transmission losses of 3\% ($\alpha=0.97)$ and 5\% ($\alpha=0.95$) 
are shown in Figs.~\ref{fig:5} and \ref{fig:7}, respectively. 
Comparing Fig.~\ref{fig:1} and Fig.~\ref{fig:3}, we see that, as 
can be expected, the presence of amplification loss has an adverse 
effect on the EPR-like entanglement that can be observed at the output; 
that is, all the power spectra are amplified at frequencies 
up to $10^7$ rad/s. 
Figs.~\ref{fig:3}-\ref{fig:7} then show that the presence of 
increasing transmission losses leads to a corresponding increase 
in the power spectra across the same frequencies and thus 
a worsening of the quality of the EPR-like entanglement in the output fields. 
However, in all of these cases, it is clear that the presence 
of a controller leads to an improvement in the entanglement. 
Moreover, this is despite the fact that the controller was 
designed under the assumption that $\chi=0$ and $\alpha=1$. 
That is, the controller exhibits a level of robustness 
in its ability to improve the system performance. 

\begin{figure}[tbph]
\centering
\includegraphics[scale=0.3]{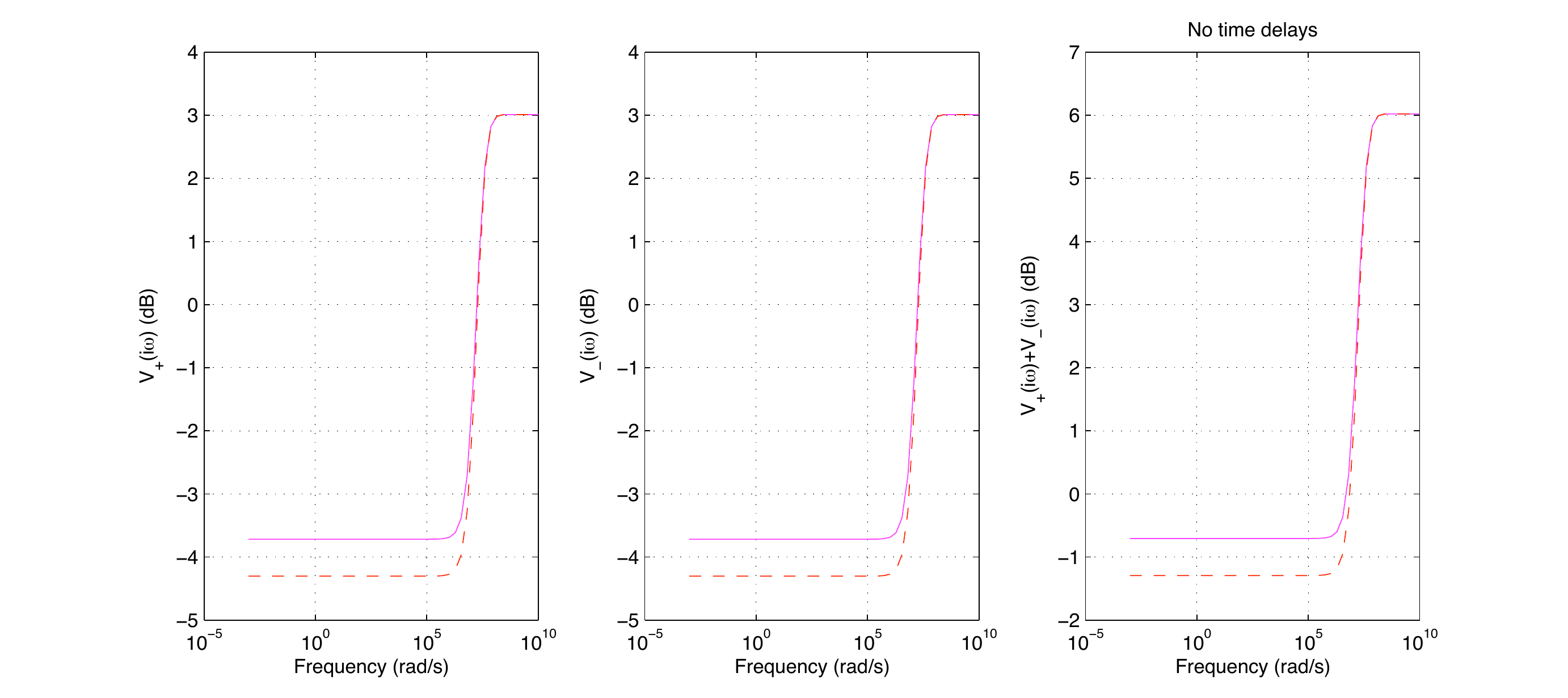}
\caption{Performance of the system when the amplification loss 
coefficient is $\chi=5.5902 \times 10^6$ Hz, transmission losses of 5\%, 
and there are no time delays. 
}
\label{fig:19}
\end{figure}

\begin{figure}[tbph]
\centering
\includegraphics[scale=0.3]{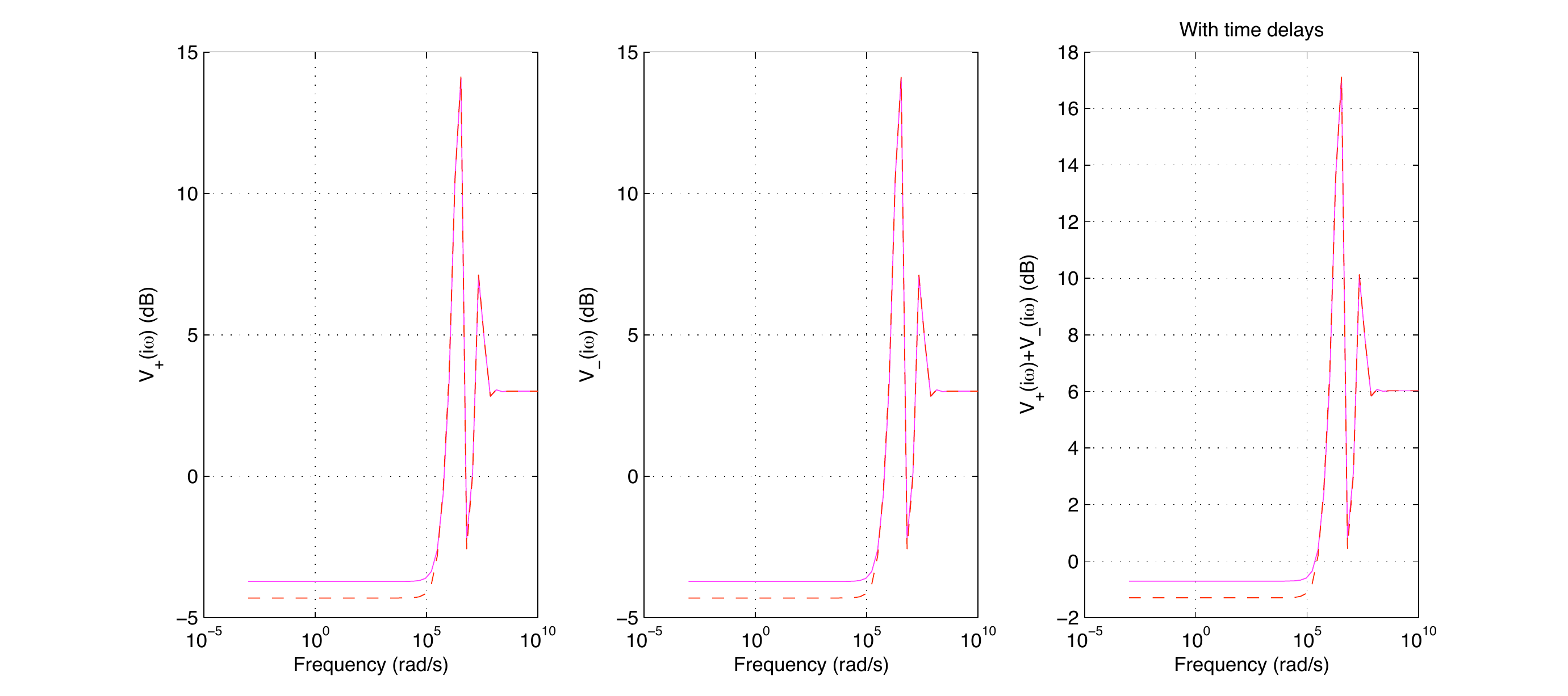}
\caption{Performance of the system when the amplification loss 
coefficient is $\chi=5.5902  \times 10^6$ Hz, transmission losses of 5\%, 
and there are time delays $T=10^{-6}$~s and $T_m=2 \times 10^{-6}$~s. 
}
\label{fig:20}
\end{figure}

Finally, Fig.~\ref{fig:19} shows the case where 
$\chi=5.5902 \times 10^6$ Hz, which is a substantial percentage of 
amplification loss compared to pump intensity, and the transmission loss is $5\%$. 
Here, again, the time delays are assumed to be zero. 
The figure indicates that  the LQG controller 
still manages to improve entanglement although the power 
spectra are, as can be expected, noticeably higher than those 
in Figs.~\ref{fig:1}-\ref{fig:7}. 
Finally, for the same values of $\chi$ and $\alpha$, the effect 
of the presence of the time delays $T= 10^{-6}$~s and 
$T_m=2 \times 10^{-6}$~s is shown in Fig.~\ref{fig:20}. 
As we had seen in Fig.~\ref{fig:2}, the delays reduce the frequency 
range in which a reduction in the power spectra by the controller can be observed. 
In particular, in the frequency range up to about $10^{5}$ rad/s, 
where the reduction is observed, the amount of reduction is 
roughly the same as what can be achieved without the delays. 
Also, again using DDE-BIFTOOL, we can inspect that the system is 
stable (i.e., the symmetrized covariance matrix of the system converges as $t \rightarrow \infty$). 
Also, although not shown here, we remark that in general the controlled 
system is able to remain stable even when the time delays $T$ and 
$T_m$ are increased up to 0.1 and 0.2 seconds, respectively. 
However, of course, it is undesirable to work in such severe cases 
of time delays,  since it means that the system
reaches steady state rather slowly. 

We conclude by remarking that  in the non-quantum setting there are sophisticated methods for designing controllers beyond the LQG that specifically take into account time delays (see, e.g., \cite{Moelja2006} and the references cited therein ), which have the potential to be adapted to the quantum setting. These more advanced controllers may potentially offer further improvements to the power spectral profile but are, however, beyond the scope of the present paper. Time delays are important in the quantum network setting where nodes can have dynamical time scales that are shorter  than the time scales for  propagation of fields between nodes. Thus control of quantum networks with time delays is a timely research topic since this scenario would be encountered in many quantum networks of interest. To the best of the authors' knowledge, only a few papers have so far considered quantum feedback control in the presence of time delays, e.g., \cite{Nishio2009,Emary2012}.

\section{Conclusion}
\label{sec:conclusion} 

This paper has developed and studied a distributed 
entanglement generation scheme for two continuous-mode output 
Gaussian fields that are radiated by two spatially separated 
Gaussian oscillator systems. 
It is shown that a LQG measurement-feedback controller can be 
designed to enhance the EPR-like entanglement between 
the two output fields across a certain frequency range, even in the 
presence of important practical 
imperfections in the system.
It is demonstrated that the controller displays a degree of 
robustness in the sense that although it was designed 
for the ideal scenario it can still provide an enhancement and stability over 
the case of  no controller being present, despite the presence of 
the imperfections. In summary, the results reported here indicate the potential 
utility of feedback controllers in the task of distributed 
entanglement generation using distributed resources. \\

\section*{ACKNOWLEDGEMENTS}
H.N. acknowledges the support of the Australian Research Council and the Japan Society for the Promotion of Science (JSPS). N.Y. wishes to acknowledge the support of JSPS Grant-in-Aid No. 40513289.


\begin{thebibliography}{}


\bibitem{Cirac1997}
J. I. Cirac, P. Zoller, H. J. Kimble, and H. Mabuchi, 
Phys. Rev. Lett. 78, 3221 (1997).

\bibitem{Kimble2007}
C. W. Chou, J. Laurat, H. Deng, K. S. Choi, H. de Riedmatten, 
D. Felinto, and H. J. Kimble, 
Science 316, 1316 (2007).

\bibitem{Bennett1996}
C. H. Bennett, H. J. Bernstein, S. Popescu, and B. Schumacher, 
Phys. Rev. A 53, 2046 (1996). 

\bibitem{Ferraro2005}
A. Ferraro, S. Olivares, and M. G. A. Paris, 
e-print arXiv:quant-ph/0503237 (2005). 

\bibitem{Eisert2002}
J. Eisert, S. Scheel, and M. B. Plenio, 
Phys. Rev. Lett. 89, 137903 (2002).

\bibitem{Mancini2007}
S. Mancini and H. M. Wiseman, 
Phys. Rev. A 75, 012330 (2007).


\bibitem{Serafini2010}
A. Serafini and S. Mancini, 
Phys. Rev. Lett. 104, 220501 (2010).


\bibitem{Carvalho2008}
A. R. R. Carvalho, A. J. S. Reid, and J. J. Hope, 
Phys. Rev. A 78, 012334 (2008).

\bibitem{Yamamoto2008}
N. Yamamoto, H. I. Nurdin, M. R. James, and I. R. Petersen, 
Phys. Rev. A 78, 042339 (2008).

\bibitem{WisemanBook}
H. M. Wiseman and G. J. Milburn, 
{\it Quantum Measurement and Control}, 
(Cambridge University Press, 2010). 

\bibitem{Yan2011}
Z. Yan, X. Jia, C. Xie, and K. Peng, 
Phys. Rev. A 84, 062304 (2011).

\bibitem{Doherty1999}
A. C. Doherty and K. Jacobs, 
Phys. Rev. A 60, 2700 (1999). 

\bibitem{Belavkin2008}
V. P. Belavkin and S. C. Edwards, 
Quantum filtering and optimal control, 
in Quantum Stochastics and Information: 
Statistics, Filtering and Control, 143-205, 
(World Scientific, 2008). 

\bibitem{GardinerBook}
C. W. Gardiner and P. Zoller, 
{\it Quantum Noise}, 
(Springer-Verlag, Berlin and New York, 3rd edition, 2004).

\bibitem{Ou1992}
Z. Y. Ou, S. F . Pereira, and H. J. Kimble, 
Appl. Phys. B 55, 265 (1992).

\bibitem{Vitali2006}
D. Vitali, G. Morigi, and J. Eschner, 
Phys. Rev. A 74, 053814 (2006). 



\bibitem{Iida2012}
S. Iida, M. Yukawa, H. Yonezawa, N. Yamamoto, and A. Furusawa, 
IEEE Trans. Automat. Contr. 57(8), 2045-2050 (2012).

\bibitem{Nurdin2009}
H. I. Nurdin, M. R. James, and A. C. Doherty, 
SIAM J. Control Optim. 48-4, 2686-2718 (2009). 

\bibitem{Gough2009a}
J. Gough and M. R. James, 
IEEE Trans. Automat. Contr. 54-11, 2530-2544 (2009).

\bibitem{Gough2009b}
J. Gough and M. R. James, 
Comm. Math. Phys. 287, 1109-1132 (2009).

\bibitem{Michiels2007}
W. Michiels and S. I. Niculescu, 
{\it Stability and Stabilization of Time-Delay Systems}, 
(Advances in Design and Control, SIAM, 2007).


\bibitem{Engelborghs2001}
K. Engelborghs, T. Luzyanina, and G. Samaey, 
{\it DDE-BIFTOOL vol. 2.00: A Matlab package for bifurcation
analysis of delay differential equations}, 
(Technical report TW-330, Department of Computer Science, 
Katholieke Universiteit Leuven, Leuven, Belgium, 2001).

\bibitem{Engelborghs2002}
K. Engelborghs, T. Luzyanina, and D. Roose, 
ACM Trans. Math. Softw. 28(1), 1-21 (2002).


\bibitem{Cao2003}
D. Q. Cao, P. He, and K. Zhang, 
J. Math. Anal. Appl. 283, 362-374 (2003).

\bibitem{Lei2007}
J. Lei and M. C. Mackey, 
SIAM J. Appl. Math 67, 387-407 (2007).

\bibitem{Moelja2006}
A. A. Moelja, G. Meinsma, and J. Kuipers, 
IEEE Transactions Automat. Contr. 51(8),  1347-1354 (2006).

\bibitem{Nishio2009}
K. Nishio, K. Kashima, and J. Imura, 
Phys. Rev. A 79, 062105 (2009).

\bibitem{Emary2012}
C. Emary,  e-print arXiv:1207.2910 [cond-mat.mes-hall] (2012). 


\end{thebibliography}
\end{document}